\documentclass[aps,prl,notitlepage,twocolumn,superscriptaddress,showpacs]{revtex4-1}

\usepackage{color}
\usepackage{graphicx}
\usepackage{amsmath}
\usepackage{hyperref}
\usepackage{natbib}
\usepackage{dcolumn}

\newcommand{\beginsupplement}{%
        \setcounter{table}{0}
        \renewcommand{\thetable}{S\arabic{table}}%
        \setcounter{figure}{0}
        \renewcommand{\thefigure}{S\arabic{figure}}%
        \setcounter{equation}{0}
        \renewcommand{\theequation}{S\arabic{equation}}%
     }

\begin{document}

\title{Strongly correlated growth of Rydberg aggregates in a vapor cell}

\author{A. Urvoy}
\email[Electronic address: ]{a.urvoy@physik.uni-stuttgart.de}
\affiliation{5. Physikalisches Institut and Center for Integrated Quantum Science and Technology, Universit\"{a}t Stuttgart, Pfaffenwaldring 57, 70550 Stuttgart, Germany}

\author{F. Ripka}
\affiliation{5. Physikalisches Institut and Center for Integrated Quantum Science and Technology, Universit\"{a}t Stuttgart, Pfaffenwaldring 57, 70550 Stuttgart, Germany}

\author{I. Lesanovsky}
\affiliation{School of Physics and Astronomy, University of Nottingham, Nottingham, NG7 2RD, UK}

\author{D. Booth}
\affiliation{Homer L. Dodge Department of Physics and Astronomy, The University of Oklahoma, 440 West Brooks Street, Norman, Oklahoma 73019, USA}

\author{J.P. Shaffer}
\affiliation{Homer L. Dodge Department of Physics and Astronomy, The University of Oklahoma, 440 West Brooks Street, Norman, Oklahoma 73019, USA}

\author{T. Pfau}
\affiliation{5. Physikalisches Institut and Center for Integrated Quantum Science and Technology, Universit\"{a}t Stuttgart, Pfaffenwaldring 57, 70550 Stuttgart, Germany}

\author{R. L\"{o}w}
\email[Electronic address: ]{r.loew@physik.uni-stuttgart.de}
\affiliation{5. Physikalisches Institut and Center for Integrated Quantum Science and Technology, Universit\"{a}t Stuttgart, Pfaffenwaldring 57, 70550 Stuttgart, Germany}

\date{\today}

\begin{abstract}
The observation of strongly interacting many-body phenomena in atomic gases typically requires ultracold samples. Here we show that the strong interaction potentials between Rydberg atoms enable the observation of many-body effects in an atomic vapor, even at room temperature. We excite Rydberg atoms in cesium vapor and observe in real-time an out-of-equilibrium excitation dynamics that is consistent with an aggregation mechanism.
The experimental observations show qualitative and quantitative agreement with a microscopic theoretical model. Numerical simulations reveal that the strongly correlated growth of the emerging aggregates is reminiscent of soft-matter type systems.
\end{abstract}


\maketitle

Due to their exaggerated properties, Rydberg atoms find applications in various research fields ranging from cavity QED \cite{Haroche2013}, quantum information \cite{Saffman2010}, quantum optics \cite{Dudin2012,Peyronel2012,Maxwell2013,Gorniaczyk2014,Tiarks2014}, microwave sensing \cite{Sedlacek2012} to molecular physics \cite{Bendkowsky2009}. One particular area of research employs the strong interactions between Rydberg atoms to create strongly interacting many-body quantum systems for quantum simulation \cite{Weimer2010,Guenter2013,Ravets2014}, quantum phase transitions \cite{Low2009}  and the realization of correlated or spatially ordered states \cite{Pohl2010, Cinti2010, VanBijnen2011, Schauss2012, Schauss2014}. In any system, gaseous, liquid, glass or solid, spatial correlations can only arise if interactions are present. 
In Rydberg gases these correlations have recently been revealed by the direct imaging of the resonant excitation blockade effect \cite{Schauss2012,Schwarzkopf2013}. 
In our experiment an initially nearly ideal gas of thermal atoms at room temperature is excited into a  strongly interacting Rydberg state. The Rydberg excitations show correlated many-body dynamics that shares similarities with that of soft-matter systems.

\begin{figure}[t]
    \includegraphics[scale=1]{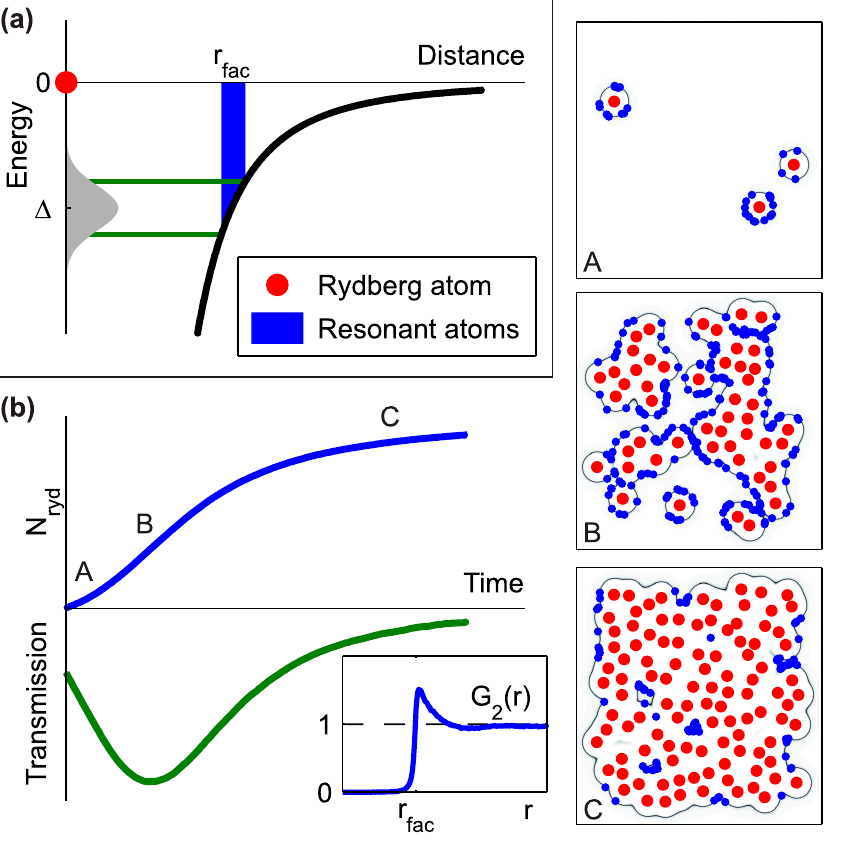}
    \caption{Principle of the aggregation. (a) Interaction induced level shift. An already excited Rydberg atom (red dot) at the origin produces an energy shift for the neighboring atoms. At the facilitation radius $r_{\rm fac}$, the atoms are exactly shifted in resonance with the excitation laser, red-detuned by $\Delta$. The grey shaded area symbolizes the excitation bandwidth, a Gaussian whose width is given by the dephasing rate $\gamma$. (b) Left: Typical time evolution of the Rydberg population and of the transmission change of the exciting laser [experimental signal, which is proportional to the derivative of the Rydberg density, see text and Fig.~\ref{fig:fig2} for details] during the aggregation process. 
    Right: Snapshots of the aggregation at times A, B and C, illustrated for a frozen gas in 2D. Red dots correspond to Rydberg atoms. Blue dots correspond to resonant atoms. The dark grey line shows the `resonant shell' in which the excitation of Rydberg atoms is facilitated. 
    Inset: Sketch of the density-density correlation function expected for the steady state [corresponding to (C)]. 
    }
    \label{fig:fig1}
\end{figure}

We consider the situation of a dense atomic gas, where dense means that the mean inter-particle distance is smaller than the typical interaction distance. 
The interplay between these strong interactions and an off-resonant excitation laser results in a non-equilibrium dynamics with pronounced spatial correlations 
(related is the antiblockade in ultracold atoms \cite{Ates2007,Amthor2010}).
At a certain distance $r_{\rm fac}$ from a Rydberg atom the interaction shifts the Rydberg states in resonance with the detuned excitation light fields, as depicted in Fig.~\ref{fig:fig1}(a).
The subsequent resonant excitations are restricted to specific distances and therefore result in the formation of so-called aggregates that feature non-trivial spatial correlations \cite{Lesanovsky2014, *Marcuzzi2014, Garttner2013, Schempp2014, Malossi2014}. 
This so-called facilitated excitation occurs predominantly at the boundary of aggregates [see Fig.~\ref{fig:fig1}(b)], similarly to seeded nucleation. The number of Rydberg excitations grows increasingly faster as the size of the aggregates gets larger. After a certain time a (quasi) steady state is reached and the excitation saturates because the sample is maximally filled with Rydberg atoms. 
The spatial correlations expected for this steady state are depicted in Fig.~\ref{fig:fig1}(b). The density-density correlation function sketched here shows an enhanced probability for finding two atoms separated by the facilitation radius $r_{\rm fac}$. There is no apparent long-range order since the isotropic facilitation mechanism does not favor the creation of crystalline arrangements.
The existence of such Rydberg aggregates has recently been observed in ultracold systems \cite{Schempp2014, Malossi2014}. In these studies the Rydberg number distributions were extracted and matched to the results of empirical rate equation models, proving the existence of small aggregates with up to 8 excitations. 

In this Letter we investigate a complementary regime and study the dynamics of the aggregation process in an atomic vapor at room temperature. 
Central parameters such as the dephasing rate and the atomic density are up to 2 orders of magnitude larger than in previous work \cite{Schempp2014, Malossi2014}. 
To quantify the non-equilibrium dynamics we perform a systematic study of the characteristic time scale that underlies the aggregation process. We find that the functional dependence of this time scale on various system parameters can be approximated by power laws. The corresponding exponents agree well with those obtained from a theoretical aggregation model. Further evidence for the aggregation dynamics is found through a quantitative evaluation of the number of Rydberg excitations and the resulting mean inter-particle distance which is consistent with the existence of a facilitation radius.

\begin{figure}[t!]
    \includegraphics[scale=1]{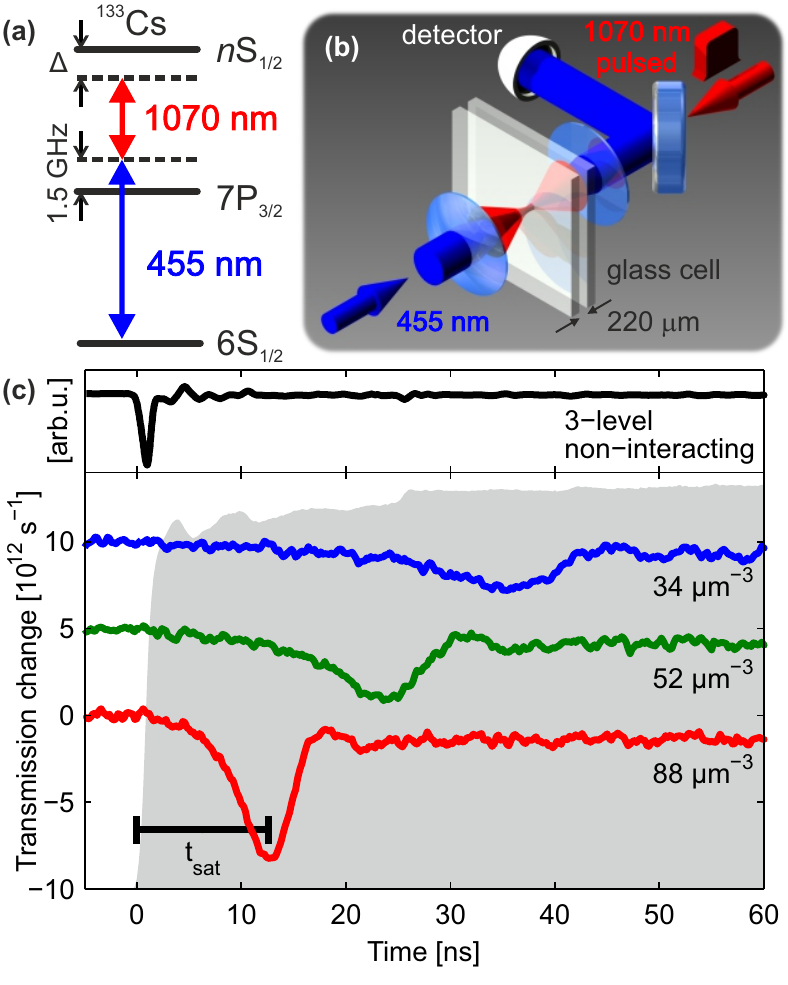}
    \caption{(a) Schematic energy level diagram for the excitation to Rydberg states of cesium atoms. (b) Sketch of the experimental setup. (c) 
    Transmission signals at different atom number densities for the 32S state. The two-photon detuning is $ \Delta \approx 2\pi \times -2200 $~MHz. The two-photon Rabi frequency is $ \Omega = 2\pi \times 100 $~MHz. The grey-shaded area in the background represents the temporal pulse shape of the infrared laser. The time delay $t_{\rm sat}$ used for the quantitative analysis is also depicted. The black curve (upper panel) is the expected transmission signal (not to scale) for non-interacting atoms (following the approach from Ref. \cite{Huber2011}).
    }
    \label{fig:fig2}
\end{figure}

We perform our experiments in a gas of cesium atoms at room temperature or above. The cesium atoms are confined in a 220~${\rm \mu{}m}$ thick glass cell [see Fig.~\ref{fig:fig2}(b)].  The atomic density $N_{\rm g}$ ranges from $10$~${\rm \mu{}m}^{-3}$ to $500$~${\rm \mu{}m}^{-3}$, which is large compared to previous experiments on interacting Rydberg atoms in thermal vapor \cite{Raimond1981, Baluktsian2013, Carr2013} and non-BEC cold atomic gases \cite{Schempp2014, Malossi2014, Heidemann2007} where $N_{\rm g} \lesssim 10$~${\rm \mu{}m}^{-3}$.
We excite cesium atoms from the ground state ${\rm 6S_{1/2}}$ to a Rydberg state $n{\rm S_{1/2}} $ off-resonantly via the intermediate state ${\rm 7P_{3/2}}$ [see Fig.~\ref{fig:fig2}(a)]. The two-photon transition is red-detuned by $\Delta$ with respect to the unperturbed Rydberg state. The lower transition is driven by a cw laser at 455~nm, which also serves as the probing field. For the upper transition a laser at $\sim 1070$~nm is pulsed on the nanosecond scale using a fast Pockels cell. The effective Rabi frequency of the two-photon transition is then given by $\Omega=\Omega_{455}\, \Omega_{1070}/(2\, \delta_{\rm 7P})$, where $\Omega_{455}$ and $\Omega_{1070}$ are the respective single transition Rabi frequencies and $\delta_{\rm 7P}=1.5$~GHz is the detuning to the intermediate state \cite{SuppMat}. 
We measure the transmission change of the lower 455~nm laser after suddenly switching on the upper 1070~nm laser on. Note that this signal is directly proportional to the time derivative of the Rydberg population \cite{SuppMat}.
Typical transmission traces are shown in Fig.~\ref{fig:fig2}(c) for various atom number densities. Immediately after the 1070~nm laser is switched on (at $\rm t=0$) atoms are transferred to the Rydberg state, and for each excited Rydberg atom one photon is removed from the probe resulting in a decrease of the transmission.
One striking feature is the large red detuning of up to $10$~GHz at which these signals are observed, in contrast to previous work on coherent excitation in a more dilute gas \cite{Huber2011, Baluktsian2013}. Line broadenings and shifts to the blue have been observed with S-states in cold atoms \cite{Schempp2014}, and Rydberg aggregates have been observed only at blue detunings \cite{Schempp2014, Malossi2014}, as expected from the repulsive Rydberg-Rydberg interaction potential for most Rydberg S-states \cite{Singer2005}. However at large atomic densities, the mean inter-particle distance is so small that the perturbative character of the van-der-Waals potential is not valid anymore. Level crossings between pair states now have to be considered. For Rydberg S-states in cesium a strong resonant dipole-quadrupole interaction with neighboring pair-states occurs at $0.55$~${\rm \mu{}m}$ for $n=32$ \cite{SuppMat,Schwettmann2006}. Initially dipole-forbidden attractive pair states acquire some \mbox{$n$S-$n$S} admixture at short inter-atomic distances due to the dipole-quadrupole interaction. Therefore the Rydberg-Rydberg interaction for S-states in cesium also features an effective attractive component, allowing for us to observe an excitation signal at large red detunings.

A first indication for the collective nature of the excitation dynamics is provided by the data in Fig.~\ref{fig:fig2}. Here we consider the characteristic time scale of the excitation $t_{\rm sat}$, i.e. the time at which the transmission reaches its minimum [as shown in Fig.~\ref{fig:fig2}(c)]. In the non-interacting case this time scale is given by $ 1/\Delta \sim 0.1$~ns (black line). In the experiment however the excitation signal is slow ($t_{\rm sat} \gg 1/\Delta$) and density dependent, which is only possible if many-body effects are included. Pulse propagation effects can be excluded due to the small optical depth at the large detunings applied.

To analyze the dynamics, we systematically varied all accessible parameters independently: $\Delta$, $\Omega$, $N_{\rm g}$ and $n^*$, where $n^* = n-\delta$ is the effective principal quantum number of the Rydberg state, $\delta$ being the quantum defect. Only the dephasing rate $\gamma$, determined by the Doppler shifts, the velocity of the atoms and the steep slope of the pair-state potentials, cannot easily be tuned over a significant range (for the derivation of the actual dephasing rate, see Supplemental Material \cite{SuppMat}). The basic experimental settings are the following: $\Delta \approx 2\pi \times -2200 $~MHz, $ \Omega = 2\pi \times 100 $~MHz, $N_{\rm g}=36$~${\rm \mu{}m}^{-3}$ and $n^*=27.95$ (32S state), and are all but one kept constant. 
First, by fitting the values of the time scale of the excitation $t_{\rm sat}$ against $\Delta$ we obtain a power law as $t_{\rm sat}\propto |\Delta|^a$ with $a=1.99(14)$. In the same way, we find that $t_{\rm sat}\propto \Omega^b$ with $b=-2.10(5)$. The power law for $N_{\rm g}$ is $t_{\rm sat}\propto N_{\rm g}^c$ with $c=-1.09(6)$. Finally we performed density scans at three different principal quantum numbers (32S, 34S and 36S states) and were able to extract a scaling behavior as $ N_{\rm g}^{-1} (n^*)^d$ with $d=-4.6(5)$. These results are summarized in Table~\ref{tab:tab1}.

By integrating the transmission signal over time we obtain an estimate of the actual number of Rydberg atoms present in the sample. At the time when the Rydberg excitation starts to saturate the Rydberg number in the excitation volume reaches approximately 50000~atoms. The corresponding mean inter-atomic distance is $\langle r \rangle=0.65$~${\rm \mu{}m}$, leading to approximative values for the effective van-der-Waals or dipole-dipole strength of interaction as $C_{\alpha} = \Delta \times \langle r \rangle ^{\alpha}$ \cite{Lesanovsky2014, *Marcuzzi2014} with $C_6 = 2\pi \times -109$~$\rm MHz\cdot\mu{}m^6$ and $C_3 = 2\pi \times -405$~$\rm MHz\cdot\mu{}m^3$ (at $\Delta \approx 2\pi \times -1500 $~MHz). 
We use these values as an input for simulations described in the following. 

\begin{table}
	\begin{ruledtabular}
	\begin{tabular}{ l d d d d }
				&	\Delta\footnote{Range: -4.5 \dots -1.3~GHz}	&	\Omega\footnote{Range: 75 \dots 175~MHz}	&	N_{\rm g}\footnote{Range: 28 \dots 101~${\rm \mu{}m}^{-3}$}	&	n^*\footnote{Range: 32S - 34S - 36S}		\\ \hline 
		Exp		&	1.99(14)	&	-2.10(5)	&	-1.09(6)	&	-4.6(5)		\\ 
		vdW		&	2.15(1)		&	-2			&	-0.86(1)	&	-4.72(5)	\\ 
		dd		&	2.38(0)		&	-2			&	-0.74(1)	&	-2.95(2)	\\ 
	\end{tabular}
	\caption{ Power laws for the aggregation delay $t_{\rm sat}$. In the experiment (`Exp'), the parameters are \mbox{$\Delta \approx 2\pi \times -2200 $~MHz}, $ \Omega = 2\pi \times 100 $~MHz, $N_{\rm g}=36$~${\rm \mu{}m}^{-3}$ and $n=32$, or varied over the specified range. The power laws labeled as `vdW' and `dd' were extracted from simulations in an ensemble of randomly and uniformly distributed atoms with van-der-Waals (ensemble of $10^3$~atoms) dipole-dipole interaction (ensemble of $15^3$~atoms) respectively, using the parameters from the experiment and the extracted interaction strengths (see text). The dephasing rate was also extracted from the experiment \cite{SuppMat}. All uncertainties represent the confidence region of the fits. No uncertainty is given for the exponent of $\Omega$ in the theory, fixed at $-2$ as by the assumption of strong dephasing.}
	\label{tab:tab1}
 	\end{ruledtabular}
\end{table}

We compared these findings to the results of the Rydberg aggregation model adapted from Ref. \cite{Lesanovsky2014, *Marcuzzi2014}. This model describes the dynamics of Rydberg aggregation within a system of two-level atoms in the presence of strong dephasing. This central assumption of strong dephasing applies to the experimental situation discussed here since the motion-induced dephasing is larger than the Rabi frequency ($\gamma \gtrsim 2 \pi \times 0.5$~GHz \cite{SuppMat}). The reduction to a two-level system is well justified in our case, although the commonly used adiabatic elimination of the intermediate state typically breaks down for quantitative considerations of the coherent evolution \cite{Huber2011, Baluktsian2013}. Here however, because of the large dephasing rates due to the atomic motion we do not observe coherent evolution, and because of the large detunings we expect only less than $1.5\%$ population of the intermediate state. 
We performed simulations for an ensemble of randomly distributed atoms in 3D, assuming a pure van-der-Waals ($V_{\rm vdW}= C_6/r^6$) or pure dipole-dipole ($V_{\rm dd}= C_3/r^3$) attractive interaction potential. 
In fact the actual experimental potential curves have a more complex spatial dependence, due to strong state mixing \cite{SuppMat}. The latter makes the actual microscopic description of the excitation dynamics rather intricate since a two-level description of an atom might not be sufficient for covering all possible pair excitation channels. 
Therefore the van-der-Waals and dipole-dipole potential that are used in our two-level model should be regarded as the limiting approximations of the actual experimental situation. We evaluate the time scale of the excitation $t_{\rm sat}$ defined by the transmission minimum, as in the experiment. 
All results are listed in Table~\ref{tab:tab1}. We find an excellent agreement between the experimental and theoretical exponents. In particular the dependence of $t_{\rm sat}$ on approximately $\Omega^{-2}$ 
indicates that a strong dephasing effect is present in our system \cite{Lesanovsky2014, *Marcuzzi2014}. Moreover the power laws vary only slightly, independently of whether pure van-der-Waals or pure dipole-dipole interaction are used. Our experimental results are compatible with both potentials, suggesting a rather weak dependence on the actual shape of the potential and that the basic mechanism of aggregation only relies on the existence of a facilitation radius. 

\begin{figure}
    \includegraphics[scale=1]{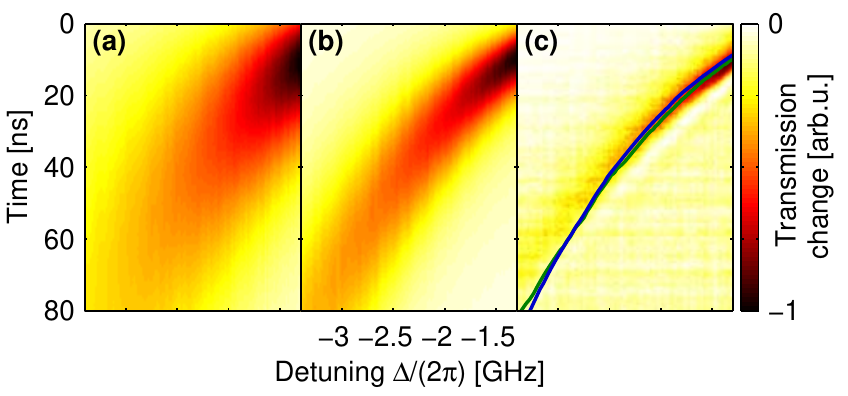}
    \caption{Density plots showing the transmission change as a function of time (vertical axis, from top to bottom) and detuning. The parameters are as in Table~\ref{tab:tab1}. (a) Simulated data with pure van-der-Waals interaction. The Rabi frequency was rescaled with a factor $0.43$. (b) Simulated data with pure dipole-dipole interaction (for $15^3$ atoms). The Rabi frequency was rescaled with a factor $0.26$. (c) Experimental data. The blue (resp. green) line shows the position of the absorption maximum $t_{\rm sat}$ in the simulated data with van-der-Waals (resp. dipole-dipole) interaction.}
    \label{fig:fig3}
\end{figure}

In Fig.~\ref{fig:fig3}(a) [resp.~(b)] we show the transmission change extracted from the model with van-der-Waals [resp.~dipole-dipole] interaction and in Fig.~\ref{fig:fig3}(c) from the experiment. The quantitative agreement for the time scale $t_{\rm sat}$ between theory and experiment is excellent. Due to the complex pair state potentials and strong state-mixing that are crucial here, an adaptation of the Rabi frequency has to be performed in order to use the two-level model (i.e.~a model with a single Rydberg state). In our simulations the Rabi frequency is reduced by a factor of 0.38 (for van-der-Waals interaction) and 0.23 (for dipole-dipole interaction) compared to the experiment. These factors are compatible with the $n$S-$n$S admixture that one can extract from the pair-state potentials \cite{SuppMat}. Note that 
with 
dipole-dipole interaction there is a dependence on the system size due to the long-range character of the $1/r^3$ potential, hence the different rescaling factors applied in Fig.~\ref{fig:fig3}(a) and \ref{fig:fig3}(b).

The comparison of the experimental results to those of the model show that facilitated excitation is occurring, which implies that an aggregation dynamics takes place. The frozen gas approximation which is used to model the aggregation in \cite{Lesanovsky2014} holds strictly for time intervals of about 0.3~ns, the transit time of the atoms through the `resonant shell' (see Figure~\ref{fig:fig1}). During this time aggregates consisting of up to 4 atoms are formed \cite{SuppMat}. Their spatial correlations will then quickly decay to those of a non-interacting ideal gas. Note, however, that even after that time the aggregation of excitations continues at the boundary of the uncorrelated aggregates. 

Despite this effect the agreement between the experimental values of $t_{\rm sat}$ and those obtained from the simple theoretical model is in general good [see Fig.~\ref{fig:fig3}]. Yet the time evolution differs as the experimental signal appears much more narrow in time than in the simulations. We attribute this discrepancy to the inherent differences between the microscopics of the experimental system and the theoretical model (e.g. the complex potentials and the atomic motion). Interestingly, when considering the temporal width of the signal, the simulations with dipole-dipole interaction seem to agree best with the experiment, which is realistic considering that one component of the actual interaction potentials is of pure dipole-dipole nature.
Moreover, the model does not account for the fact that the excitation of isolated, i.e. not facilitated, atoms is in fact coherent. 

Another feature of the experimental data that is not reproduced by the theory is the small overshoot in transmission just before the system reaches the steady state [for instance at 20~ns for $N_{\rm g} = 88$~${\rm \mu{}m}^{-3}$ in Fig.~\ref{fig:fig2}(a)]. Remnants of coherent Rabi oscillations can be ruled out given the large dephasing rates and large detunings with respect to the single-atom resonance. Moreover the time scale of the overshoot, e.g. the time between the two maxima, does not show a specific dependency on the detuning, as is the case for Rabi flopping. Explaining this feature will require further study and a more sophisticated theory.
Note that ion effects can be neglected. 
There are no ions in the sample prior to the excitation, otherwise they would act as aggregation seeds \cite{Ions} at $t=0$ and change the signal dramatically. Moreover plasma formation would occur only once Rydberg atoms have been excited \cite{Vitrant1982} and therefore cannot explain our Rydberg excitation signal. In spite of the low collisional ionization rate ($\sim 40$~MHz \cite{Vitrant1982}) it might happen that in the course of the experiment single ions are created. However single ions will not disturb the aggregation dynamics, because of the similarity between the ion-Rydberg and the Rydberg-Rydberg potentials \cite{Ions}.

In conclusion, we have shown a facilitated excitation process of Rydberg atoms in thermal vapor. We have characterized the dynamics by measuring power laws for all relevant experimental parameters. These power laws, as well as absolute numbers for the dynamics and Rydberg density are in excellent agreement with those from a model for Rydberg aggregation. 
These results were obtained in the framework of thermal vapors, but the general ideas also apply to experiments in ultracold atoms. 
In the ultracold regime it would be interesting to study how quantum effects influence the aggregation time scale when the strength of dephasing noise is systematically lowered. 

\begin{acknowledgments}
The authors would like to thank D.~Peter, S.~Weber, S.~Hofferberth and J.P.~Garrahan for fruitful discussions, as well as H.~K\"{u}bler for proof reading. 
This project was supported by the Carl-Zeiss-Stiftung and the ERC under contract number 267100. 
I.L. acknowledges funding from the European Research Council under the European Union's Seventh Framework Programme (FP/2007-2013) / ERC Grant Agreement n. 335266 (ESCQUMA) and the EU-FET Grant No. 512862 (HAIRS). 
D.B. and J.P.S. acknowledge financial support from grants NSF(PHY-1205392) and DARPA(60181-PH-DRP) for this work. 
\end{acknowledgments}

\bibliographystyle{apsrev4-1}

\begin{thebibliography}{36}
\bibitem [{Ion()}]{Ions}%
  \BibitemOpen
  \href@noop {} {\bibinfo  {journal} {The Stark-shift of an ion on the 32S
  state is $C_4/r^4$ with $C_4 = 2\pi \times -213$~$\rm MHz\cdot\mu{}m^4$,
  which is an attractive potential with comparable strength to the
  Rydberg-Rydberg potentials. Therefore facilitation would similarly occur
  around an ion, if created}\ }\BibitemShut {NoStop}%
\end{thebibliography}

%


\section*{Supplemental Material}
\beginsupplement

\subsection*{Experimental details}
The 455~nm laser is frequency-stabilized with a blue detuning of $\delta_{7{\rm P}}=1.5$~GHz to the $6{\rm S}_{1/2}, {\rm F}=4 \rightarrow 7{\rm P}_{3/2}, {\rm F'}=5 $ transition (see Fig.~\ref{fig:fig2}). The 1070~nm laser is scanned over the two-photon resonance to the Rydberg state $|n{\rm S}_{1/2} \rangle$. 
Both lasers have an estimated linewidth below 5~MHz. 
The frequency of the 1070~nm laser is calibrated for each measurement using a Fabry-P\'{e}rot interferometer and additionally by an EIT-signal \cite{Urvoy2013} to fix the origin of the frequency axis. The 455~nm laser typically has a power of 3~mW. The 1070~nm laser has a power of 15~W and passes through a Pockels cell, allowing to switch the power of this laser for the experiment as fast as $1.5$~ns with a repetition rate of 10~kHz. 
The two-photon Rabi frequency is $ \Omega = 2 \pi \times 0.05 \dots 0.5 $~GHz, the detuning to the Rydberg state is $ \Delta = 2 \pi \times - 1 \ldots -10 $~GHz. 

The glass cell is home-made and consists of two 1~mm-thick quartz optical flats of $5\times5$~${\rm cm}^2$ separated by a 220~${\rm \mu{}m}$ spacer and sealed at the edge. A glass tube was connected to the cell, filled with cesium under vacuum and sealed off. This glass tube serves as a reservoir. The temperature of the cell is kept constant at $200^\circ$C to prevent the metallic cesium from condensing, whereas the temperature of the reservoir is varied between $70^\circ$C and $150^\circ$C to tune the atom number density. 
The atom number density in the cell is determined for each measurement by performing absorption spectroscopy on the D2-line of cesium \cite{Siddons2008}. 
Both beams are linearly polarized, overlapped in a counter-propagating configuration and focused inside the cell to a beam waist of approx. 15~${\rm \mu{}m}$. After passing through the cell the blue beam is focused on a pinhole in order to select the central part of radius 6.25~${\rm \mu{}m}$ inside the cell. We verified that the resulting imaged volume is indeed almost cylindrical inside the cell. The change in transmission is detected using a fast amplified silicon photodiode (Femto HSA-X-S-1G4-Si). 
During the measurement, the transmission change of the 455~nm laser is monitored and averaged 300~times by a fast oscilloscope.
We ensured that we operate in the linear response regime of the photodiode and measured the conversion efficiency of the detection to be approx. 850~V/W. This allows us to convert the transmission change into the real number of Rydberg atoms that are excited, assuming that no decays and losses of Rydberg atoms occur. This assumption is valid for short times as the time spent by an atom inside the excitation volume is on the order of 20~ns.

\subsection*{Approximation to 2-level atoms}

\begin{figure}
    \includegraphics[scale=1]{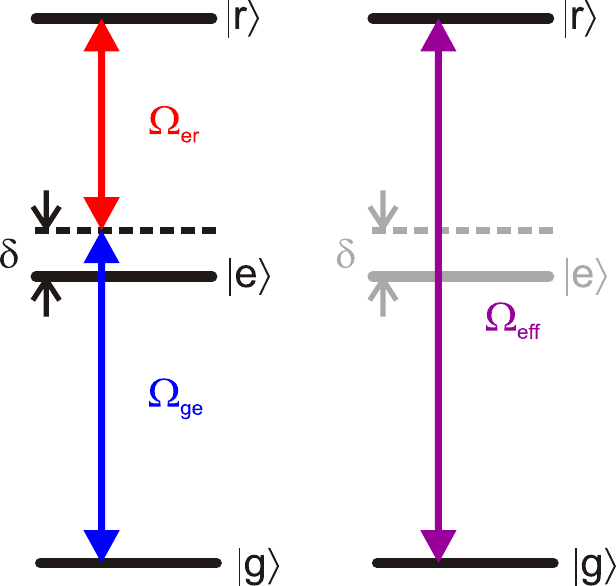}
    \caption{Left panel: 3-level ladder system with the useful definitions. The transition between the levels $|i\rangle$ and $|j\rangle$ is driven with the Rabi frequency $\Omega_{ij}$. The detuning to the intermediate state is $\delta$. Right panel: Approximation to a 2-level system after adiabatic elimination of the intermediate state $|{\rm e}\rangle$. The effective 2-level Rabi frequency is $\Omega_{\rm eff} = \Omega_{\rm ge}\Omega_{{\rm er}}/(2|\delta|)$. For simplicity reasons the detuning to state $|{\rm r}\rangle$ and the arising light shifts are not shown.}
    \label{fig:figS1}
\end{figure}

We reduce our 3-level system to a 2-level system by making use of the adiabatic elimination of the intermediate state. The basic principles of this approximation and the relevant parameters are shown in Fig.~\ref{fig:figS1}. The density matrix for the 3-level system is $\rho=(\rho_{ij})_{i,j=\{{\rm g,e,r}\}}$, and we define the density matrix for the effective 2-level system as $\hat{\rho}=(\hat{\rho}_{ij})_{i,j=\{{\rm g,r}\}}$. The populations of states $|{\rm g}\rangle$ and $|{\rm r}\rangle$ coincide between both descriptions, i.e. $ \rho_{{\rm gg}} = \hat{\rho}_{{\rm gg}} $ and $ \rho_{{\rm rr}} = \hat{\rho}_{{\rm rr}} $. The approximation relies on assuming that the intermediate state remains unpopulated ($\partial_t \rho_{{\rm ee}}=0$ and $\left.\rho_{{\rm ee}}\right|_{t=0}=0$). 
The resulting identity from the optical Bloch equations without decays is
\begin{equation}
	0 = \Omega_{{\rm ge}} \operatorname{Im}(\rho_{{\rm ge}}) - \Omega_{{\rm er}} \operatorname{Im}(\rho_{{\rm er}})
\end{equation}
Using $\partial_t \rho_{{\rm rr}} = \Omega_{{\rm er}} \operatorname{Im}(\rho_{{\rm er}})$ one obtains
\begin{equation}
	\operatorname{Im}(\rho_{{\rm ge}}) = \frac{1}{\Omega_{{\rm ge}}} \partial_t \rho_{{\rm rr}}
\end{equation}
which means that the absorption on the lower transition is proportional to the excitation rate to the Rydberg state $|{\rm r}\rangle$. For the 2-level system we also obtain from the optical Bloch equations 
$\partial_t \hat{\rho}_{{\rm rr}} = \Omega_{\rm eff} \operatorname{Im}(\hat{\rho}_{{\rm gr}})$ 
and therefore 
\begin{equation}
	\operatorname{Im}(\rho_{{\rm ge}}) = \frac{\Omega_{\rm eff}}{\Omega_{{\rm ge}}} \operatorname{Im}(\hat{\rho}_{{\rm gr}}) 
								 	   = \frac{\Omega_{{\rm er}}}{2|\delta|} \operatorname{Im}(\hat{\rho}_{{\rm gr}}) 
\end{equation}
This last relation shows the link between the 2-level model and what is experimentally measured.

\subsection*{S-state potentials in cesium}

\begin{figure}
    \includegraphics[scale=1]{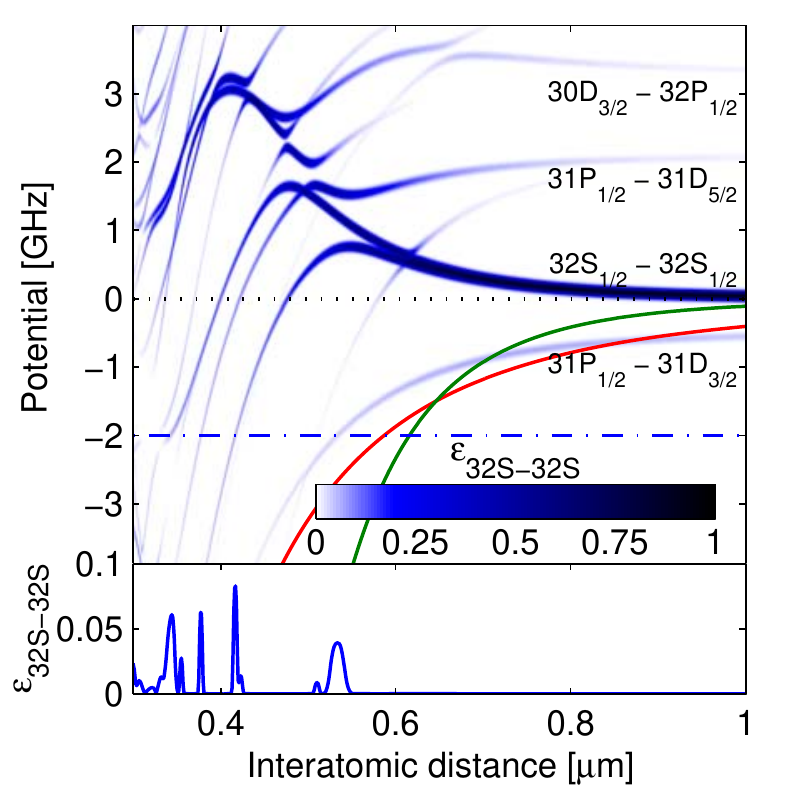}
    \caption{Top panel: Density plot of the 32S-32S admixture $\varepsilon_{\rm 32S\text{-}32S}$ versus the inter-atomic distance. The molecular quantum number here is $M=0$, and interactions up to the quadrupole-quadrupole order are included \cite{Schwettmann2006}. The green (resp. red) line depicts the extrapolated van-der-Waals (resp. dipole-dipole) pair-state potential. Bottom panel: 32S-32S admixture $\varepsilon_{\rm 32S\text{-}32S}$ versus the inter-atomic distance at a $-2$~GHz potential energy (depicted as the blue dash-dot line in the upper panel).}
    \label{fig:figS2}
\end{figure}

At short inter-atomic distances, i.e. below 1~${\rm \mu{}m}$, the pair-state interaction potentials for Rydberg S-states become very complex, as shown in Fig.~\ref{fig:figS2} for the 32S state. We will focus on this 32S state, but the situation is similar for other principal quantum numbers. Neighboring $n^{\prime}{\rm P}\text{-}n^{\prime\prime}{\rm D}$ pair-states interact with the $\rm 32S\text{-}32S$ state with weak but resonant dipole-quadrupole interaction, leading to avoided crossings and state mixing. This means that the $n^{\prime}{\rm P}\text{-}n^{\prime\prime}{\rm D}$ pair-states, which are dipole-forbidden for laser excitation from the $7{\rm P}_{3/2}$ state in the non-interacting case, carry some admixture $\varepsilon_{\rm 32S\text{-}32S}$ of the $\rm 32S\text{-}32S$ state. Therefore it is possible to excite pair-states at negative detunings, where the detuning is defined with respect to the unperturbed $\rm 32S\text{-}32S$ state, which would not be the case with purely repulsive van-der-Waals interaction. 

These pair state potentials are actually only valid for an excitation close to an atom in the 32S state. Once such a pair is created, it is mainly a $n^{\prime}{\rm P}\text{-}n^{\prime\prime}{\rm D}$ pair, with an amplitude of $1-\varepsilon_{\rm 32S\text{-}32S}^2 > 0.99$. For any subsequent excitation around this pair, one therefore has to consider the pair-state interaction potentials for the $32{\rm S}\text{-}n^{\prime}{\rm P}$ and $32{\rm S}\text{-}n^{\prime\prime}{\rm D}$ pair states. For an S-P pair state the interaction potential is purely of the form $C_3 / r^3$, since the dipole-dipole interaction of an S-P pair with its permutation P-S is resonant, and the symmetric (resp. antisymmetric) linear combination exhibits attractive (resp. repulsive) interaction. The corresponding interaction strength is $C_3^{\rm S\text{-}P} \approx -290$~$\rm MHz\cdot\mu{}m^3$, which is in excellent agreement with the value of $C_3 = 2\pi \times -258$~$\rm MHz\cdot\mu{}m^3$ extracted from the experimental results. For an S-D pair state the interaction is of van-der-Waals character and repulsive ($C_6^{\rm S\text{-}D} \approx 11$~$\rm MHz\cdot\mu{}m^6$). The $n^{\prime\prime}{\rm D}$ component therefore plays no role in our case of aggregation with red detunings.

The admixture $\varepsilon_{\rm 32S\text{-}32S}$ is the rescaling factor for the Rabi frequency when exciting a  $n^{\prime}{\rm P}\text{-}n^{\prime\prime}{\rm D}$ pair. Because of the nature of the aggregates, which enclose previously excited Rydberg atoms, at most every second Rydberg excitation is created via this process. The other excitations occur either as a direct off-resonant excitation from the ground state, or close to a $n^{\prime}{\rm P}\text{-}n^{\prime\prime}{\rm D}$ pair. The Rabi frequency does not need rescaling in both cases. An estimate for the overall rescaling factor is therefore $\sqrt{\varepsilon_{\rm 32S\text{-}32S}} \sim 0.3$, which is close to the rescaling factors of 0.38 (for van-der-Waals interaction) and 0.23 (for dipole-dipole interaction) needed to match the experimental data to the results from the simulation.

In the whole treatment, we neglected the distance dependency of $\varepsilon_{\rm 32S\text{-}32S}$. In order to justify this approximation, let us consider just two pair states $|1\rangle$ and $|2\rangle$ with energies $E_1(r) = C_6/r^6$ and $E_2(r) = E_0 - C_6/r^6$, and a small dipole quadrupole interaction $W(r) = U/r^4$ between the two pair states. The resulting admixture of pair state $|1\rangle$ at small $r$ is $\varepsilon_1 \approx W(r)/|E_2(r)-E_1(r)| \propto r^{2}$ if we consider $E_0$ to be small, and the pair state energy is $E_1^{\prime}(r) = \Delta \approx -C_6/r^6$. Therefore $\varepsilon_1 \propto \Delta^{1/3} $ under these rather crude approximations. Using the same argument as before we obtain $\Omega \propto \Delta^{1/6}$ which is a sufficiently weak dependence to be neglected. 

\subsection*{Dephasing rate}
The first contribution to dephasing in the experiment is the Doppler effect, characterized by the two-photon Doppler width \mbox{$\gamma_{\rm D} = |k_{455}-k_{1070}| \sqrt{\frac{8 \ln2\ k_B T}{m}}$}. Here $m$ is the atomic mass of cesium, $T$ is the temperature and $k_{455}$ and $k_{1070}$ the wavenumbers of the two lasers, from which we consider the difference because the lasers are counter-propagating. At $200^\circ$C we have $\gamma_{\rm D} = 2 \pi \times 512 $~MHz.

\begin{figure}[t]
    \includegraphics[scale=1]{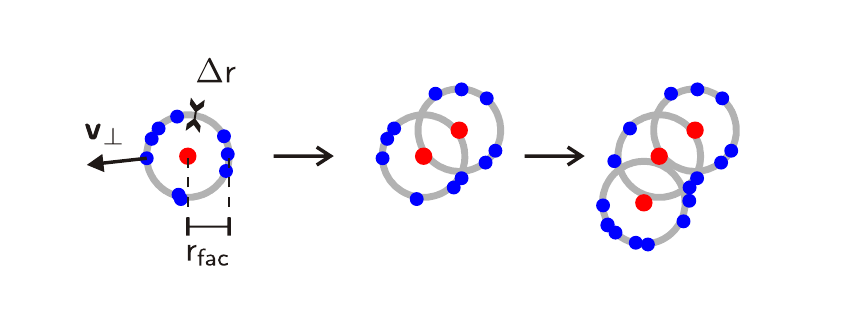}
	\begin{ruledtabular}
	\begin{tabular}{ l c c c c c }
		$n$					&	1	&	2	&	3	&	4	&	5	\\ 	\hline
		$P_n(t_{\rm f})$	&	0.91	&	0.084	&	0.0061	&	0.00039	&	0.000023	\\ 
		$\eta_n$ 			&	4740	&	436		&	32		&	2.0		&	0.12		
	\end{tabular}
 	\end{ruledtabular}
    \caption{Top: Sketched snapshots of the excitation process of a 3-atom aggregate. Red (resp. blue) dots represent Rydberg (resp. ground state) atoms. The grey lines show the resonant shell of width $\Delta r$ for each atom, within which the excitation is facilitated. The velocity component ${\bf v}_{\perp}$ perpendicular to the resonant shell is shown for an atom inside the shell. 
    Bottom: Probabilities $P_n(t_{\rm f})$ of creating an $n$-atom aggregate in an interval $t_{\rm f}$ and typical number of $n$-atom aggregates $\eta_n$ for $N_{\rm fac} = 5000$~facilitating atoms, similarly to the experimental situation.}
    \label{fig:figS3}
\end{figure}

The other important source of dephasing arises from the short transit time during which the atoms are in the resonant shell (or facilitation region) of width $\Delta r$. In first approximation, only the velocity component perpendicular to the shell $\textbf{\textsf{v}}_{\perp}$ determines the transit time, as depicted in Fig.\ref{fig:figS3}. Therefore we define this motional dephasing rate as  
\begin{equation}
	\gamma_{\rm m} = \frac{\langle \textsf{v}_{\perp} \rangle}{\Delta r} \label{eq:eqS4}
\end{equation}
where $\langle \textsf{v}_{\perp} \rangle = \frac{1}{\sqrt{3}} \sqrt{\frac{8 k_B T}{\pi m}}$ is the one dimensional mean atomic velocity. 
For a temperature of $200^\circ$C, this 1D mean velocity is $\langle \textsf{v}_{\perp} \rangle = 158$~${\rm m.s^{-1}}$.
Since both dephasing mechanisms are gaussian, the total dephasing rate is defined as $\gamma = \sqrt{(\gamma_{\rm D})^2 + (\gamma_{\rm m})^2 }$. 

Moreover $\Delta r$ is related to the total dephasing rate and on the interaction potential (see Fig.1).
First assuming Rydberg interaction of the van-der-Waals (vdW) type, we rename the relevant quantities as $\Delta r^{\rm vdW}$, $\gamma^{\rm vdW}_{\rm m}$ (motional dephasing rate) and $\gamma^{\rm vdW}$ (total dephasing rate). The width of the resonant shell is also given by \cite{Lesanovsky2014, *Marcuzzi2014}
\begin{equation}
	\Delta r^{\rm vdW}	\approx \frac{1}{3} \left( \frac{C_6}{\Delta}\right) ^{\frac{1}{6}} \left( \frac{\gamma^{\rm vdW}}{2 |\Delta|} \right) \label{eq:eqS5}
\end{equation}
By combining equations \eqref{eq:eqS4} and \eqref{eq:eqS5} we obtain the following self consistent equation for the motional dephasing rate $\gamma^{\rm vdW}_{\rm m}$:
\begin{equation}
	\frac{\langle \textsf{v}_{\perp} \rangle}{\gamma^{\rm vdW}_{\rm m}}	= \frac{1}{3} \left( \frac{C_6}{\Delta}\right) ^{\frac{1}{6}} \left( \frac{\sqrt{\left(\gamma_{\rm D}\right)^2 + \left(\gamma^{\rm vdW}_{\rm m}\right)^2 }}{2 |\Delta|} \right)  \label{eq:eqS6}
\end{equation}
Solving equation \eqref{eq:eqS6} yields \mbox{$\gamma^{\rm vdW}_{\rm m}= 2 \pi \times 493$~MHz}, so that the total dephasing rate is \mbox{$\gamma^{\rm vdW} = 2 \pi \times 711$~MHz}. 

For the case of pure dipole-dipole (dd superscript) interaction, equation \eqref{eq:eqS5} changes to
\begin{equation}
	\Delta r^{\rm dd}	\approx \frac{2}{3} \left( \frac{C_3}{\Delta}\right) ^{\frac{1}{3}} \left( \frac{\gamma^{\rm dd}}{2 |\Delta|} \right) 
\end{equation}
and the dephasing rate can be estimated to \mbox{$\gamma^{\rm dd} = 2 \pi \times 591$~MHz} with \mbox{$\gamma^{\rm dd}_{\rm m} = 2 \pi \times 296$~MHz}. 
$\gamma^{\rm vdW}$ and $\gamma^{\rm dd}$ are the values that were used in the simulations.

\subsection*{Aggregate size}

Since we are not in the frozen gas regime, aggregates are only spatially correlated like the ones shown in the inset of Fig.~1(b) for a short period of time. At longer times the atomic motion will destroy the spatial correlations between the atoms. This will, however, have no impact on the fact that Rydberg excitations are facilitated by already existing one. In the following we will estimate the size of the 'frozen aggregates' [Fig.~1(b)]. The time scale over which the ensemble can be considered frozen is given the transit time of the atoms through the resonant shell of width $\Delta r$, given by $t_{\rm f} = (\gamma_{\rm m})^{-1} \approx 0.32$~ns.
During this time interval the ensemble can be considered as a frozen gas for our purposes. 
We estimate the size of the aggregates that can be formed during this time in a very simplified model. The parameters are the same as in Fig.~2(c) with the extracted values of $\gamma$ and $C_6$ : $\Delta \approx 2\pi \times -2200 $~MHz, $ \Omega = 2\pi \times 0.4 \times 100 $~MHz, $N_{\rm g}=88$~${\rm \mu{}m}^{-3}$, $C_6 = 2\pi \times -109$~$\rm MHz\cdot\mu{}m^6$ and $\gamma^{\rm vdW} = 2 \pi \times 711$~MHz. $0.4$ is the rescaling factor for $\Omega$ discussed in the main text. 

Let us first consider one individual Rydberg atom which acts as the seed for an aggregate. In the resonant shell (sphere of radius $r_{\rm fac} = 0.61$~${\rm \mu{}m}$ and width $\Delta r = 51$~nm, see Eq.~\eqref{eq:eqS4}) around this first atom reside $\nu_{\rm res} = N_{\rm g} \times 4\pi r_{\rm fac}^2 \Delta r = 20.7$~atoms in average.
Moreover the size of the resonant shell grows with the number of Rydberg atoms. 
For this we approximate an $n$-atom aggregate to a sphere (whose volume is occupied by $n$~Rydberg atoms separated by the facilitation radius). Then the entire resonant shell contains $n^{2/3}\times \nu_{\rm res}$ atoms. 
For each ground state atom in the shell, the time constant for the excitation is \mbox{$\tau_0 = \gamma/\Omega^2 \approx 71$~ns}, and therefore the excitation rate of the $(n+1)$-th atom is $\Gamma_{n+1} = n^{2/3} \times \nu_{\rm res}/\tau_0 $. 
If we simplify the problem and consider that the atoms are excited sequentially, the probabilities $P_n(t)$ that an aggregate formed by $n$ Rydberg atoms at the time $t$ per pre-existing Rydberg atom obeys the following rate equation:
\begin{align}
	\partial_n P_n 	&= \Gamma_{n-1} P_{n-1} - \Gamma_{n} P_{n}  \nonumber \\
					&= \frac{\nu_{\rm res}}{\tau_0} \left( (n-1)^{2/3} P_{n-1} - n^{2/3} P_{n} \right)
\end{align}
In the table in Figure~\ref{fig:figS3} we show the solutions of these equations for aggregates consisting of up to 5 atoms at the time $t_{\rm f}$.  

The number of $n$-atom aggregates that are excited in the whole excitation volume during $t_{\rm f}$ is given by $\eta_n = N_{\rm fac} \times P_n(t_{\rm f})$, where $N_{\rm fac}$ is the number of Rydberg atoms which can act as a nucleation grains. 
As the excitation occur at the boundary of the already excited Rydberg ensemble $N_{\rm fac}$ is only a fraction of the $\sim 50000$ Rydberg atoms. 
The values of $\eta_n$ are shown in Figure~\ref{fig:figS3}, with $N_{\rm fac} = 5000$ (corresponding to the outer shell of a sphere with $50000$ Rydberg atoms). 
For this Rydberg atom number the largest spatially correlated aggregates that we create in our experiment can thus be estimated to consist of approximately 4 atoms.

\end{document}